\documentclass[11pt]{article}
\usepackage{graphicx}

\begin{document}

\title{Density distribution configuration and development of vortical patterns in accreting  close binary star system}

\author{Daniela Boneva, Lachezar Filipov\\
Space Research and Technology Institute,
 Bulgarian Academy of Sciences,\\
 \newline 
danvasan@space.bas.bg, lfilipov@space.bas.bg }
\maketitle

\begin{abstract}
We investigate the problem of structures formation in accretion disc zone, resulting from the tidally interaction in close binary star system. We aim to examine the area where the incoming flow meets the matter around secondary star and the resulting effects throughout the accretion disc. 
The  research is based on employment of the fluid dynamics equations in conjunction with numerical simulations leading to the design of graphical models of accretion processes.  For the simulation we propose box-framed sharing schemes. 
(i) The tidal transfer of matter through the inner Lagrangian point in close binary stars disturbs the flow in disc’s zones and outer disc area. Calculations on the perturbed parameters of density and velocity reveal formation of a thickened zone  in the contact area of interacted flows. It appears to be stable for a number of periods and is unaffected by rotation and dissipation processes. (ii) The results also show development of undulations, which grow to vortical patterns in the accretion disc's zone. It is confirmed that under the influence of tidal wave the conditions of reconfiguring of accretion flow structure are generated.
\end{abstract}

{Keywords}: Accretion; accretion disks; Hydrodynamics; Waves; Methods: numerical; (Stars): binaries: close;

\section{Introduction}

Study of the flow structure in astrophysics is important, because the results could be used both for consideration of the evolution status of binary stars and for the interpretation of observational data. Therefore, when investigating close components, it is necessary to include physical essence of the flow dynamics response to the interaction processes. In the astrophysics of binary stars, one of the most investigated phenomena are vortexes. Vortices and spiral configurations play a key role in the accretion disc dynamics, because they are considered as an efficient mechanism of angular momentum transportation (Baranco \& Marcus 2005) in regions where the magneto-rotational instability (Balbus \& Hawly 1998) does not operate. At present, there are many hydrodynamics studies in finding the way of vortices appear and behave in the flow. Li et al. (2000, 2001) have shown that vortices are formed by Rossby waves instability and
moving radially and thus transport mass through the disc. Vortices can be generated by a globally unstable radial entropy
gradient (Klahr \& Bodenheimer 2003) that may result in local outward transport of angular momentum. In astrophysics, the
problems of structures development have been investigated mainly numerically
(e.g. Lithwick 2009; Godon \& Livio 1999). Bracco et al. (1998) by using two-dimensional, incompressible fluid dynamics, show that anticyclonic vortices shift out and that smaller vortices merge to form larger vortices. Godon \& Livio (1999, 2000) confirm this result with two-dimensional, compressible, barotropic simulations. 
  Shen et al. (2006) examine the formation of 2D vortices starting from 2D turbulence in fully compressible
simulations. Barranco \& Marcus (2005) compute the evolution of 3D
vortices and show that part of the vortical formations could be destroyed, but the other part survive for several hundreds of orbits. They use an "inelastic code" with vertical stratification and examine the influence of 3D perturbations over the vortices in the middle disc's zone. By performing a series of runs with zero initial vorticity and perturbation wavelengths, Johnson and Gammie (2006) give a very realistic way of the initial vorticity generation. They have noted that the remaining vorticity can be generated from "finite-amplitude compressive perturbations". Johnson and Gammie
(2005) argue that the vortices are "long-lived" assuming they
could not be conserved in 3D calculations. Lesur and
Papaloizou (2009) have obtained that most anticyclonic vortices are unstable. Their
simulations show that only the vortex core is stable and that the
instability appears at the vortex boundary. Formation of vortices could be accompaned with spirals and vice-versa. Fridman (2007) confirms that for one arm spiral waves in the disc, based on the
spiral wave theory. Sawada, in his paper (Sawada et al. 1986), finds that a two-armed spiral structure developed in the result of 2-D simulations of mass
transfer via Roche lobe overflow. Their research
has been followed by a series of investigations (e.g. Rozyczka \& Spruit
1993), finding that during the stage of spirals development the appearance
of vortices are observed, after the interaction event. In our
previous studies we have found that  presence of one-armed spiral structure could be arised as an effect of 
tidal interaction of two stars in binary system (Boneva 2010(1), Boneva
2010(2)). The same spiral wave structures are observed in the calculations
of (Bisikalo et al 2008; Steeghs et al. 1997). Currently, the question of how the structures in accretion discs arise is still open for discussion, along with numerous unresolved problems concerning the formation mechanism and the duration of vortices existence in discs. By applying numerical calculations on gas-dynamical flow, we suggest  modeling of patterns formation and explanation of supporting physical processes in interacting flows. The paper is organized as follows: Section 2 presents some base equations and the
methods of analysis on it. In Section 3 we introduce the main results that reveal the variations in disc's and vicinity's  density. The development of vortical structures
formation is examined and described in Section 4.

\section{Basic equations and methods}

\subsection{Basic equations}

The nature of the interaction between a flow of matter and an envelope of two star components requires employment of gas-dynamics equations. Therefore, to obtain solutions of the above stated problem a system of equations is needed. Herein, the basic equations are presented in a form that have been suggested and affirmed by many authors: (Shore 2007; Clark \& Carswell 2007; Frank et al. 2002; Graham 2001; Shu 1992). 

We present the equations in their vector form. The equation of mass conservation is:

\begin{equation}
\frac{\partial \rho }{\partial t}+\nabla .\left( \rho v\right) =0;  \label{1}
\end{equation}

The existence of viscous processes in the accretion flow, as well as influence of
forces and rotation could be performed by the following useful form of the
Navier-Stokes equations, suggested by (Thorn 2004) and simplified here:

\begin{equation}  \label{2}
\frac{\partial v}{\partial t} +v.\nabla v=-\frac{1}{\rho } \nabla P- \\ 
\Omega\times \left(\Omega \times r\right)-2\Omega \times v-\nabla \Phi +\nu \nabla
^{2} v
\end{equation}

Where the basic notations are: $\rho $ is the mass density of the
flow, $v$ - is the velocity of the flow; $P$- is the pressure; $\nu $- is
the kinematic viscosity; $\Omega $ - is the angular velocity; $\Omega
\times \left(\Omega \times r\right)$ - is the centrifugal acceleration of
the centrifugal force; and $2\Omega \times v$ - is the Coriolis acceleration
in the mean of the Coriolis force. In the current analysis $\rho \ne const$
and $\nu \ne 0$. $\Phi $ is the gravitational potential and
it depends on the density distribution inside  each of the star's component
(Boyarchuk et al. 2002). Then the gravitational potentials $\Phi _{1} $ and $%
\Phi _{2} $ could be defined from the Poisson's equations: $\Delta \Phi _{1}
=4\pi G\rho _{1} $ and $\Delta \Phi _{2} =4\pi G\rho _{2} $

The energy balance equation for a viscous non-ideal fluid is:

\begin{eqnarray}  
\frac{\partial }{\partial t} \left[\rho \left(\frac{1}{2} v^{2}   
+\varepsilon
+\Phi \right)\right]+\nabla .\left[\rho v\left(\frac{1}{2} v^{2} +h+\Phi
\right)-2\eta \sigma .v\right]=0;
\end{eqnarray}
Where $\frac{\partial }{\partial t} \left[\rho \left(\frac{1}{2} v^{2}
+\varepsilon +\Phi \right)\right]$ is the total energy density, where the
first term on the left denotes the kinetic energy, the second is the
internal energy and the third expresses again the full potential of the
gravitational fields.

And $\left[\rho v\left(\frac{1}{2} v^{2} +h+\Phi \right)\right] $ is the
total energy flux, where $h=\varepsilon +P/\rho $ is the enthalpy, $\eta $
is the shear (or dynamical) viscosity of the flow, and $\sigma $ is the rate
of shear.

The equation of state for compressible flow is:

\begin{equation}
P=c_{s}^{2}\rho  \label{4}
\end{equation}

where $c_{s}$is the sound speed.

We present the equations in the above system in their common form and
we can easily transform them into quantities for each of the posted problems.

Next to explore is the vortical transport equation, because it is related to the examination of transfer's mechanisms in the flow. This equation could be derived in the following commonly used way, how it has been done by Nauta (2000), Lithwick (2007),
Godon (1997). If the curl of Navier-Stokes equations is considered, and use
the expressions: $\Psi =\nabla \times v$ - expressing the vorticity in the
flow; $\left( v.\nabla \right) v=\nabla \frac{v^{2}}{2}-(v\times \Psi )$ and 
$\left( \frac{\partial \Psi }{\partial t}+v.\nabla \right) \frac{1}{\rho }=%
\frac{\nabla \rho \times \nabla P}{\rho ^{3}}$; then the following
expression is obtained:

\begin{equation}  \label{5}
\frac{\partial \Psi }{\partial t} +\Psi \left(\nabla
.v\right)+\left(v.\nabla \right)\Psi = \\
-\frac{\nabla p\times \nabla \rho }{
\rho ^{2} } +D\nabla ^{2} \Psi
\end{equation}

Here $\Psi $- is the vorticity; $D$- is the diffusion coefficient (or matrix
of the transport coefficient).

This equation expresses the relation between the transport coefficient,
which takes part in the angular momentum transfer, evolution of the
vorticity with time and the non-conserve relationship between density and
pressure in the flow. Non-conservancy of specific vorticity by the each
fluid element is observed in the right-hand side of Eq. (\ref{5}), which
gives a condition for baroclinicity in the flow (Lovelace et all. 1999), and
enables the development of vortices in the accretion matter. The
baroclinicity of the general flow is given by the baroclinic term (Klahr
2004; Petersen 2007): $\nabla \rho \left( r,z,\varphi \right) \times \nabla
p\left( r,z,\varphi \right) \neq 0$. The importance for this instability is
the non-axisymetric deviations from the mean state which can lead to the
rise of the baroclinic term even in two dimensions: $\nabla \rho \left(
r,\varphi \right) \times \nabla p\left( r,\varphi \right) \neq 0$ and
vorticity can be generated. This instability, known as baroclinic (Klahr \&
Bodenhiemer 2000, 2003) and arising vortices are discussed in Section 4 of
the current survey.

\subsection{Numerical methods}

The complexity of the studied problem and analysis on the corresponding equations in hydrodynamical matter, require applying of numerical codes. We use "PDEtools" and "PDEsolve" packages, implicated in Maple 5 and Maple 12 (Maple Tutorials of Monnagan et al. 1998; Heal et al. 1998). We chose to insert into calculations the methods, which are employed in these codes, known as:
the Runge-Kutta (implicit part) method (further referred to as RK),
Alternating direction implicit method (ADI),
CenteredTime1Space(CTS)[forward/backward], BackwardTime1Space
(BTS)[forward/backward], based on finite difference scheme or Godunov's
algorithm. More detail description for RK and ADI methods could be found in (Autar \& Egwu 2008; Chang et al. 1991) and in Maple Tutorials again for CTS and BTS. All they are implicit methods, which are
general in their application. It is suitable to use them in the solutions of
partial differential equations, because of their high stability. We give
here a short description of their operation.

The method of Runge-Kutta is widely used in physical calculations. The
method treats every step in a sequence of steps in identical
manner. That fact makes it easy to add in Runge-Kutta into relatively simple
schemes (Forsythe et al. 1977). This is mathematically proper, since any
point along the trajectory of an ordinary differential equation can serve as
an initial point.

Euler method is a part of RK family methods and it's common expression is: $y_{n+1} =y_{n}
+hf\left(x_{n} ,y_{n} \right)$, which is unsymmetrical. It advances the solution through an interval $h$, from $x_{n} $ to $%
x_{n+1} =x_{n} +h$ and uses derivative information only at the
beginning of that interval. This means that the step's error is only one
power of $h$ smaller than the correction, i.e. the member $O\left(h^{2}
\right)$ is added to that expression.

The symmetrization in the error term could cancels out the first-order error
term, making the method second order. (If the error term of the method is $%
O(h^{n+1})$, a method is known as $n$-th order). Adding up the right
combination of these, we can eliminate the error terms order by order (Cash 
\& Karp 1990). That is the basic idea of the
Runge-Kutta method and we applied its schemes to equations posted in the
current survey.

It is common that numerical codes need some control during the
processing. Adaptive step-size control is used here (Cash \& Karp 1990; Hairer \& Soderling 2005),
with the purpose of achieving some predetermined correctness in the
solution with minimum computational effort. Implementation of adaptive
step-size control requires that the stepping algorithm returns information
about its performance and estimation of its truncation error.

ADI method belongs to the group of finite difference methods and follows the
idea to split the finite difference equations in two, in relation to the
derivatives in coordinates taken implicitly. The system of equations then
becomes symmetric and tridiagonal and is usually solved with tridiagonal
matrix solver.

The centered time forward/backward space method and backward time forward/backward space method  are implicit single stage methods that can be used to find solutions to PDEs containing the derivatives v,v[x],v[t],v[xt].
In the both methods, PDEs that describe right-traveling waves should use the backward method (specified as [backward]), and specify the boundary condition at the left boundary, while left-traveling waves require the forward method (specified as [forward]), and the boundary condition at the right boundary. Numerical boundary conditions are not required. The points of discretization that are used to compute the values of the grid are diferent: (t+k,x[i]) is in BTS; (t+k/2,x[i]) in CTS.

We test the equations with the "pdetest" tool, which checks if the solution is
correct. To reduce the PDEs to a simpler problem, we applied an "ansatz" tool. The specifying initial and boundary
conditions are assigned in Sects. 3 and 4 in accordance with the presented
model. 
 Box-framed scheme has been suggested and applied. We could perform the
calculations in limited regions of all disc's areas by configuring the scheme for each problem, Then, we make the calculations inside of the box, or frame with different measurement. 
We have made the code validation, applying it to the problem of shock wave distribution in a tube. The box-framed model has been applyed again.

\subsubsection{Image resolving and processing methods}

We depict most of the results by graphic simulations, which present some full 
picture of the studied processes. Then, based on the images decipherment, we have to identify the fine forms for better interpretations of the results. For the purpose of the study the high-pass: Gaussian filters, which are adequate to enhance the structure of the graphical results and to obtain spatial filtered image (Gonzalez \& Woods 1992), are applied. They can be used to enhance edges between different
regions. The Gaussian filters are part of Convolution filters which remove the low frequency
components of an image while retaining the high frequency (Haddad \& Akansu
1991, Shapiro \& Stockman 2001) local variations. This is accomplished using
a kernel with a high central value, typically surrounded by negative
weights. High pass filter with [3 x 3] kernel is used with a value of 8 for
the central pixel and values of -1 for the exterior pixels; high pass
filters can only have odd kernel dimensions. We employ this method in
Section 4.

The convolution filters produce output images in which the
brightness value at a given pixel is a function of some weighted average of
the brightness of the surrounding pixels. Convolution with the image array
returns a new, spatially filtered image (Nixon \& Aguado 2008). Different
types of filters can be generated by selecting the kernel size and values,
producing (Vandevenne 2004-2007).

We have found the above combination of methods suitable for the exploring here processes: mass transfer, interaction of the flows and concomitant structures
formation. Application of these methods and results are presented
in the next two sections. All used equations from Sect. 2.1 are
transformed further in cylindrical frame of reference ($r,\varphi ,z$). This
system allows applying 2D and 3D numerical modeling of structure of the
flowing matter. The modeling is performed in non-inertial frame of reference
and quadratic 2D - 3D set.

\section{Disturbances in the flow density and velocity. Formation of "thickened zone".} 

The gas-dynamical analysis of the flow structure in the close binary star
system has revealed tidal interaction between the out-flowing matters from
donor through the point of libration $L_{1} $ and the flow around the
accretor (Boyarchuk et al. 2002; Bisikalo et al. 2003; Frank 2008; Pringle
1985, 1992). Bisikalo et al. (2001) have shown that even a small variation
in mass transfer rate of the binary system, could disturb the equilibrium
state of the hot accretion disc. This may cause the appearance of area with
increased density, called "blob''. Using this result as a  point, we obtained a
similar dense formation, called "thickened zone''. We used the
above numerical methods in the calculations (RK \& ADI) and reported the
solution in paper Boneva (2010) firstly. The graphical result there was
obtained only for a moment of time in one period of rotation to mean an
orbital period: $\sim t_{m0} +0.25P_{p} $ (see Fig.1a and Fig. 6 in Appx.).
Where $t_{m0} $ denotes the moment of time when the mass transfer rate
begins to change and $P_{p} $ is the orbital period of the binary system. In
the current study, the simulations are performed for three additional
moments of rotational period, until one orbital period of the system is
fulfilled, see Fig. 1 (b), (c), (d): $\sim t_{m0}; +0.57P_{p}; +0.78P_{p};  
 +0.92P_{p}$. We repeated numerical runs for $\sim 4.5P_{p} $.

We apply free boundary conditions at the outer disc edge, where the density is defined to be constant: $\rho_{out}=10^{-8}\rho_{L_{1}}$, where $\rho_{L_{1}}$ is the density of the
inner Lagrangian point $L_{1}$. In the inner regions, where the mass
transfer and the interaction of streams take place, the values of the
density, as was shown, could not remain constant. A spherical form of the
location of interaction was adopted, with a free radius. The limitations of the software imposed some employment restrictions and shortening the necessary time  for these calculations, but it does not affect the final results.

The results show that the thickened zone could exists for a long
period of time and it does not change its mean characteristics over the time
of our calculations. This is close to the result, reported by
Bisikalo et al. (2001) who have showed this for several tens of orbital periods
of the binary star system. In the current paper, we stopped our examinations before
the mass transfer rate begins to subside.

\begin{figure}\label{Figure 1}
\centering
\includegraphics[width=12cm, height=10cm]{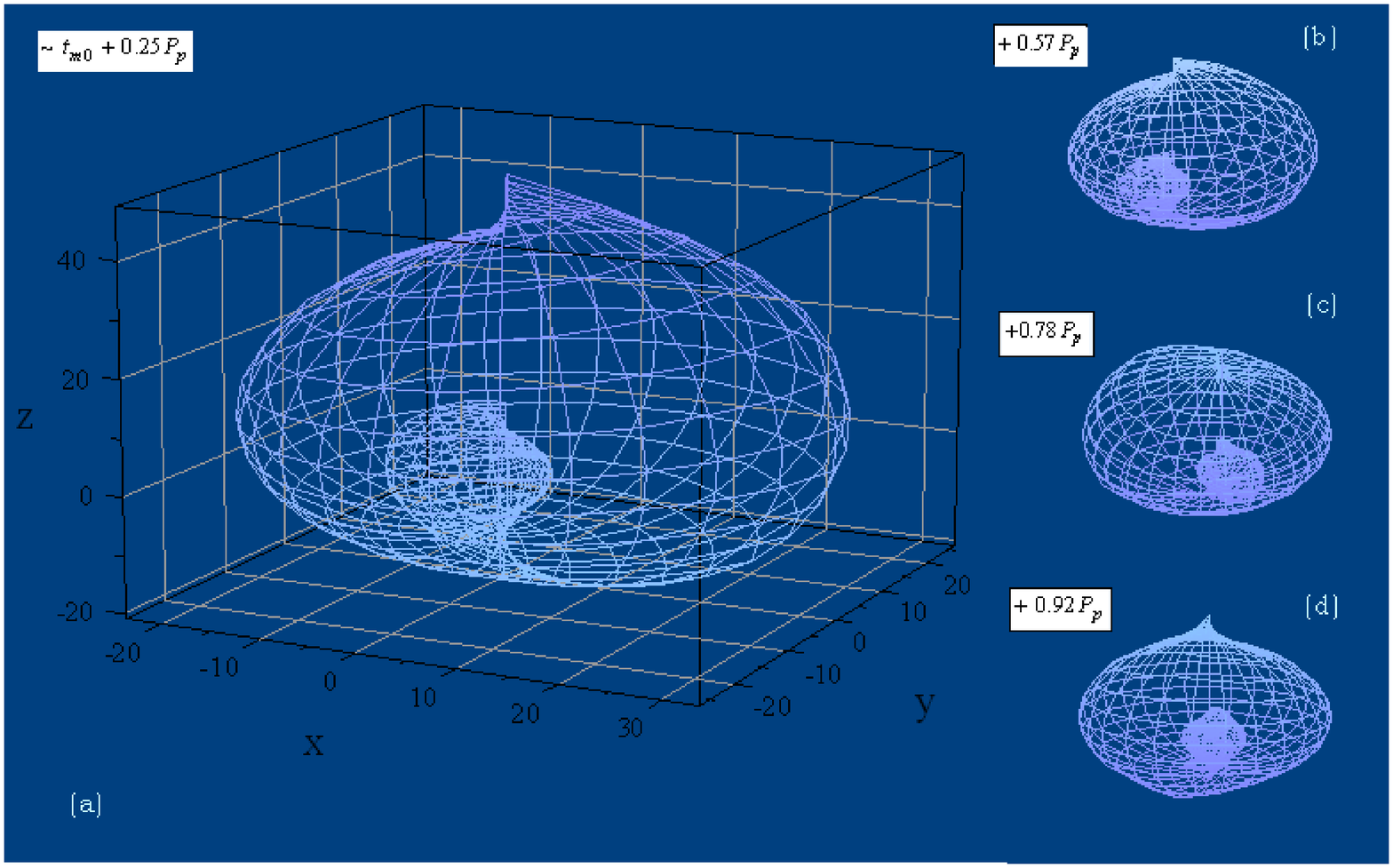}
\caption{Thickened zone formation, as a result of disturbances in the
stability state, caused by mass transfer in a binary system. The whole
system consecutive phases of rotation can be seen in four images
(a), (b), (c), (d). The dense pattern, colored in light blue, is observed in all phases at the image. The detection's stop-steps of one rotational period: $\sim t_{m0} +0.25P_{p}; \sim t_{m0} +0.57P_{p}, +0.78P_{p}, +0.92P_{p}$.The mesh grid was taken in a (x, y, z) calculation’ frame of coordinates, corresponded to the density distribution in r direction over the time t[days]. 
}
\end{figure}

In the next calculations, we can see that the changes in the mass transfer rate are in a close relation both with disturbances in
the density and in the velocity. We put values to the rate of accretion ranging: $\dot{M}_{0} \approx 10^{-10} \dot{M}_{\circ } /year\div \dot{M}_{t_{0}
+n} \approx 10^{-12} \dot{M}_{\circ } /year$. To study this, we apply the
modified perturbation function, described in Appendix and in (Boneva
2009), on the Navier-Stokes equations (Eq. 2), and obtain the ``term of
instability'': $\Omega \left(\frac{\partial r^{2} }{\partial r} \frac{V_{r} 
}{r} -\frac{\partial V_{\varphi } }{\partial \varphi } \right)$. This term
gives the relation of velocities, including perturbation value and angular
velocity. In this expression $V\left(r,\varphi \right)$ is the full quantity
of velocity, which is a sum of the perturbation value and the time averaged
velocity (see App., Eq. A1). The behavior of perturbed velocity is
calculated for different values of the dimensionless parameter $m$, which is
the mode number along $\varphi $ direction (see Appendix). After the
calculations, we detected sharp decreasing in velocity values and it is
showed in Fig.  5 (in the Appx.) and in Boneva (2009). Let's see, how the appearance of our “thickened zone” is provoked, related to the variability in density and velocity. In the places where
the velocity values are close to their minimum, the density starts to
increase and it pulls a matter there. Thereafter, this means that the
matter from a disc could be pumped out or concentrated within given places,
causing the density's dilution in close areas. We express this graphically
and the thickening place in the disc can be viewed now in form of the  pillar (see Fig.2). The result points for three periods of rotation.

\begin{figure}[htp]
\centering
\includegraphics[width=9cm, height=9cm]{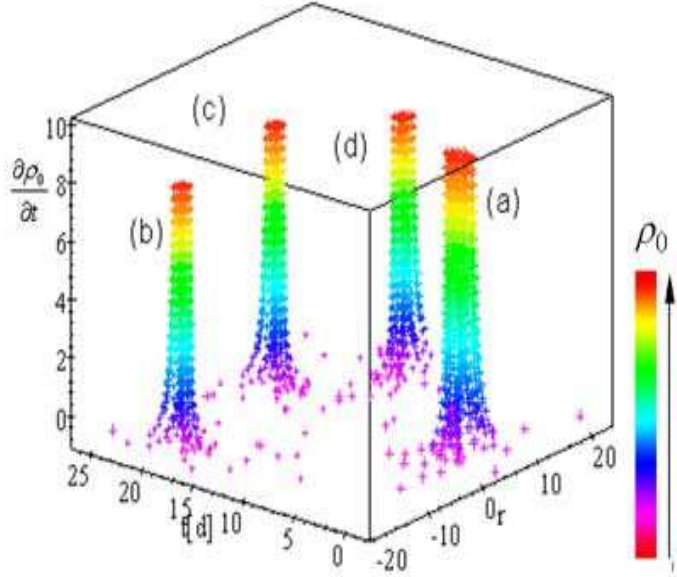}
\caption{Pillar's form graphical expression of density concentration in close area. The image shows the accumulating matter  in four moments of
time in a rotational period. Calculations are made in 3D quadratic frame in $%
x,y$ and on variations of the perturbed density value $\protect\rho^{\prime} $.}
\label{Figure 2}
\end{figure}

The pillars, seen in Fig. 2, indicate matter accumulation in the places with
increasing density, defined by the boundary conditions area, for several
periods of rotation. The image in Fig.2 corresponds to the appearance of
dense zone in Fig.1. This result is the second stage of the case presented by
Boneva (2010).

\section{Development of vortices in the disc flow.}

It was already mentioned in Sect. I and it is confirmed by  many studies vortices play a vital role in accretion disc dynamics. They are
considered to be an effective mechanism of angular momentum transport
(Barranco \& Marcus 2005). Herein, we perform a computational analysis to reveal  one possible way of their appearance by visual simulation of their development in the flow.  Our calculations are based on the vortical
transport equation (Eq. 5), because it includes the condition that could
provokes baroclinic character of the flow, (Klahr \& Bodenhiemer 2003). The box-frame model is used once again. The
introduced boundary conditions are of Dirichlet- and Cauchy type:  
$r_{v\left(1+n\right)} =K\left(x,y\right)-\frac{\partial K}{\partial r_{v} } 
\frac{\partial }{\partial t} $; $r_{v0} \left(0\right)=0$
is the radius of the vortex; 
$K\left(x,y\right)$ is the boundary area of equations activity.
We place the cylindrical coordinates $(r,\varphi ,z)$ frame for the
equations and quadratic $(x,y)$ set for the numerical scheme. We perform a series of
runs with zero initial vorticity, but different from zero the initial 
turbulence values: $v\left(0\right)=v_{0} $, $\Psi _{r,\varphi } \left(t_{0}
\right)=0$, $\rho \left(t_{0} \right)=\rho _{0} \approx 2.5\times 10^{-6}
kg/m^{-3} $, $t_{0} \approx 1$, and $r_{0} \approx 1$. Results of the simulations show  a vortex type growth in $r,\varphi $ plane of the disc zone. The box-frame values range from about $7.687\times 10{}^{-7}
AU to 6.68 \times 10{}^{-7}$ AU and from $7.687 \times\ 10{}^{-8} AU$ to $6.68
\times\ 10{}^{-8}$ AU,  corresponding to the above values of $x$
and $y$, referred to as $x_{b} $ and $y_{b} $. The development of this kind of vortices passes through three stages, resulting from the calculations. Firstly, a distortion 
of the flow laminarity is observed (Fig. 3a). In the next series of
calculations, the  velocity values step to ($v\left(t_{1} ,t_{n-1}
\right)\approx v_{0} +v_{1} $) and the layers in the examined area undergo
a weak undulation (Fig.3b). This means that the variations of velocity and
density have significant impact on the flow's behavior. For the final round
of calculations the density and velocity accepted values $\rho
\left(t_{n} \right),v\left(t_{n} \right)$ are used as an input. Then, we
consider the stage of vortex evolution in some steady period of their
development, when they are "ready'' for the angular
momentum transport (Fig. 3c). In contrast to Sect. 3, where it was shown
that the thickening patterns are formed inside the disc area, here in this
case the development of vortices is more frequently observed along the outer
sides, close to the disc's edges. In conclusion, we suppose that, when this kind of 
vortical formations leave the disc zone, they could crumble and merge into
the matter of the curcumdisc hallo, influenced by the conditions of low
density there. The solution of the last assumption is still under
investigation and the results will be presented in another report.

\begin{figure}[htp]
\centering
\includegraphics[width=15cm, height=13cm]{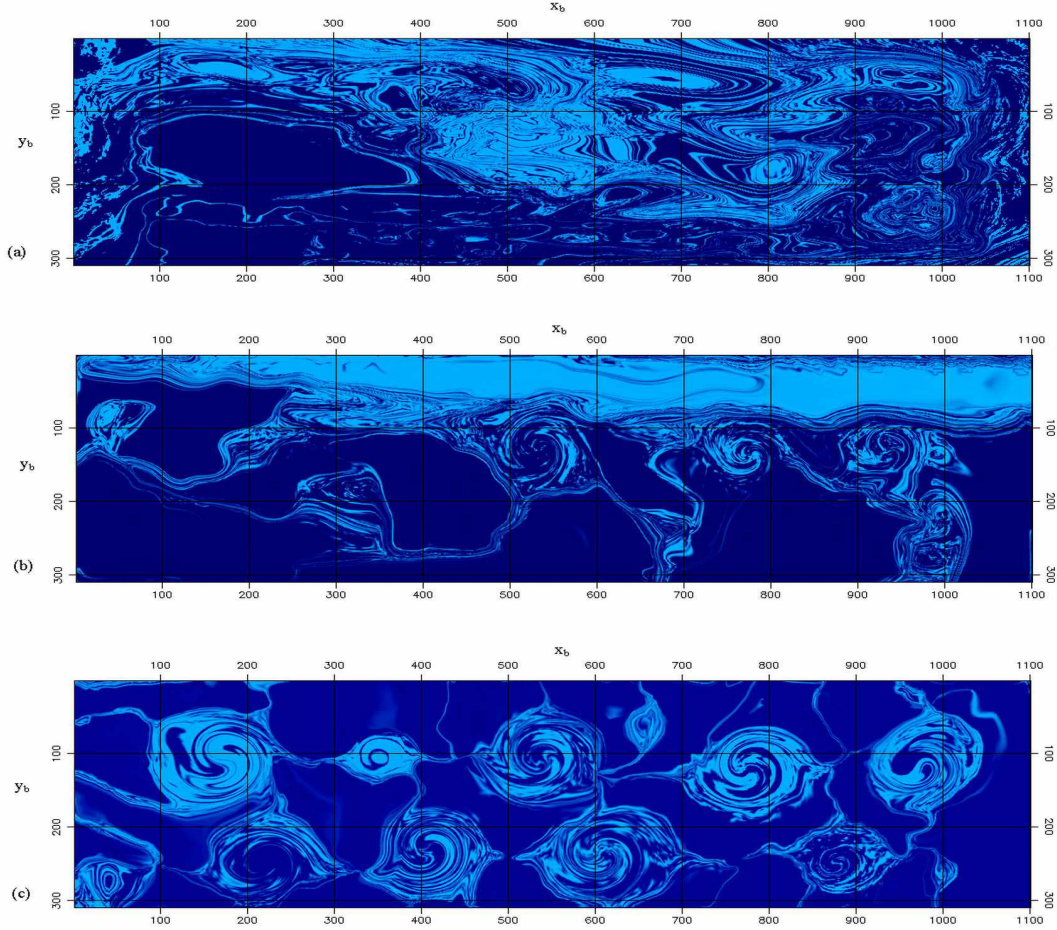}
\caption{It is seen 3 stages of the patterns development: distortion of 
laminarity of the layers (a), weakly undulations
(b), the final stage of structures formation in the flow (c). Each frame
visualizes a covered range of about $7.687 \times 10{}^{-7} AU$ to 
$6.68 \times 10{}^{-7}$ AU and $7.687 \times 10{}^{-8} AU$ to 
$6.68 \times 10{}^{-8}$ AU, reffered to the boundary frame 
$(x_{b},y_{b})$ of the calculation performance. The light Blue and 
dark Blue colours (light and dark in a grey scale for the printed 
version) show the difference in density in the
interacting flow layers. The density values are increasing from dark to
light zone.}
\label{Figure 3}
\end{figure}

We have received the view of the Fig.3 by image processing stages shown in Fig. 4. Figure 4a shows contour profile of vortex obtained
by the result of numerical simulations, which we already described in this section.  We apply on it the convolution method by high pass Gaussian filter, discussed in Sect. 2.2.2, and then depict the same vortexes in Fig.4b. The
final model is obtained by removing the outliers and eliminating the
inessential elements from the image map (Fig.4c). We made this in an attempt to better understand
the results of these simulations and distinguish the main structure
components. The image map gives a summary of the data, including the set of
projection of all examined parameters; each pixel of the image consists of
some information about the numerical graphical simulations.

\begin{figure}[htp]
\centering\includegraphics[width=15cm,height=13cm]{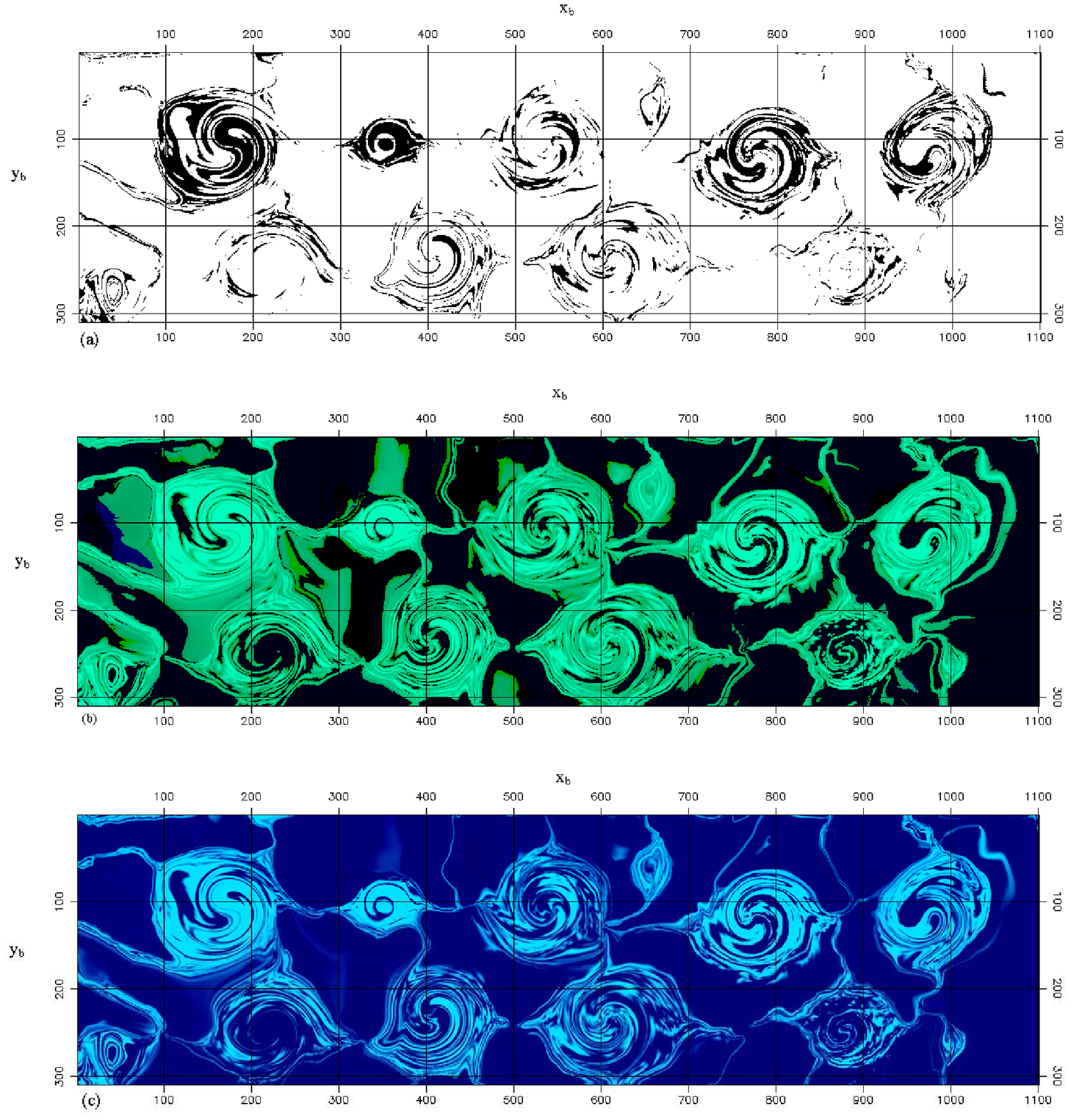}
\caption{Processing of the vortices formation stage image: (a)
Contour profile of vortices, resulting from the initial numerical
simulations on the tidal flow parameters; (b) The view of such vortexes
after applying the Convolution - high pass Gaussian filter; (c)
The final view, after removing the data outliers from the image map. }
\label{Figure 4}
\end{figure}

The baroclinicity conditions of (Klahr \& Bodenhiemer 2003),
misalignment of pressure gradient and density gradient is required wherever
there is azimuthal density gradient. This misalignment acts as a source term
for the generation of vorticity in the region of outer edge of the disc. The latter processes have initially been presented
in a similar way by Boneva \& Filipov (2009) and Boneva (2010), where the
received images  are of an isolated vortex, which is a part of the examined
region, defined above. These vortices may propagate throughout the disc,
according to the baroclinicity global character, but they are local,
temporal formations. How long these vortices would live and what would
destroy them is another unanswered question and unresolved problem. The
point we want to make is that two-dimensional simulations have been used to
identify physical processes that would likely play a role in a full
three-dimensional simulation. If vortices primarily form on the upper and
lower surfaces of accretion discs, then the radiative cooling rates could be
rapid, as was shown in the numerical
results by Barranco \& Marcus (2005). Therefore, it could be expected that
vortices confined to the surface of a disc could have longer lifetimes due
to their ability to transport heat radially outwards.

\section{Conclusion}

The investigation in this paper concerns the morphology of the flow in the
area of accretion disc during tidal interaction in the binary star.
Numerical methods were employed in the analysis of the gas-dynamical
equations, to study the processes and resulting effects, during mass
transfer between the components of the binary stars and corresponding tidal
waves. We present a modeling that reveal of how the accreting flow structure could be
 organized after interaction processes in the binary. The change of the rate in mass
 transfer causes variations in the flow density, which are expressed as accumulation of matter in some places and dilution in others. This situation leads to the appearance of the 'thickened zone' that remains stable despite of  the consequently tidal influences and dissipation processes. Besides the description in Sect. 2, the analysis show us that the construction of vortical
transport equation (Eq. 5) is some kind of the whole disc structure's
analog. It gives the relation between the angular momentum transport part
(the last term in the right hand side) and describes the sources of the
transport mechanism in sense of the vorticity function, and the
patterns developed. In the studied interaction parts of the binary star flow,
the incoming speed of flow is close to the speed of sound. Then, this tide
may causes some perturbations in quantities, which changed the dynamics of
the gas flow. After applying the numerical methods via numerical code the result
shows the presence of two-dimensional vortical patterns.
Vortices are usually local formations, but under the conditions assumed
here, they could propagate globally throughout the disc. This type of
vortices has often been observed in 2-D simulations of accretion discs.
Vortices can be generated by two-dimensional instability as the Rossby wave
instability or baroclinic instability, which have been investigated in
recent years. In this study we showed once again that under the influence of
tidal wave, the accretion flow could not remain stable and the conditions of
patterns development are generated. Then, their movement throughout
the accretion flow could support the transportation of angular momentum. The
presence of a gravity-connected companion of the star can affect the
physical processes in the star and appreciably change its evolution.
Studying the accretion disc's structure in binary stars by using
numerical methods proves that the gravitational effect of the second
component may cause the appearance of spirals shocks. Tidal waves from the donor
star, which usually cause a development of spiral density structure, could
also be responsible for vorticity in the accretion area. However the
modeling of such effects will be the topic of another report.

\section*{Appendix
}

We know from the perturbation analysis of Papaloizou \& Pringle (1984) that
in most hydrodynamical flows the flow could change its state, when some of
the parameters characterizing it (e.g. the velocity, density or pressure)
are perturbed. Then, the accretion flow could remain stable, become
instable, turbulent, or to form a vorticity. We describe here how these
perturbations are quantified and for this purpose we employ the equations of
Navier-Stockes. The first step is to introduce the perturbation quantities
into the parameters and to obtain an equation for hydrodynamic instability,
such as:

\begin{eqnarray}\nonumber
V=v+u; \rho_{0} =\rho +\rho^{\prime}; p=P+p^{\prime}; & & (A1)  
\end{eqnarray}

where $V,\rho _{0} ,p$ are the total quantities of velocity, density and
pressure, respectively; $v,\rho ,P$ are the time averaged values; $u,\rho
^{^{\prime}} ,p^{^{\prime}} $ are the perturbations in time. Balbus (2003)
has made a full and detailed analysis of the disc's stability, using similar
expression for the perturbation function, but under different conditions of
the disc flow compared to those, considered here. The perturbations are set
to be in an exponent form with a power of second order and for the velocity
they have the form: $u\left(r,\varphi \right)\approx u\exp \left(im^{2}
\varphi -i\omega t\right)$. This form is applied also to the other
quantities. Here $\omega $ is the wave number and $m$ is the mode number of $%
\varphi $ direction. For the initial values of these quantities, denoted by
``zero'' subscripts, we may write: $u_{0} \left(r,\varphi \right)\approx
u\exp \left(Q_{0} \right)$ (we apply$Q\equiv im^{2} \varphi -i\omega t$ ),
and $m_{0} \approx 1,r_{0} \approx 2,\varphi _{0} \sim \pi ,\omega t_{0}
\approx 3$. The second step is to express the Navier-Stokes equation in
cylindrical coordinates ($r,\varphi ,z$). The newly introduced quantities
from Eq.(A1) are taken into account, which give us the form of equations
transformed with perturbation values:

\begin{eqnarray}\nonumber
{\frac{\partial V_{r}}{\partial t}+V_{r}\frac{\partial V_{r}}{\partial r}+%
\frac{V_{\varphi }}{r}\frac{\partial V_{r}}{\partial \varphi }-{%
V_{\varphi }^{2}{r}}}={-\frac{1}{\rho _{0}}\frac{\partial p}{\partial r}-\Omega \frac{\partial
V_{r}}{\partial \varphi }+}2\Omega V_{\varphi }+\\ \nonumber
+{\nu \left( \frac{\partial^{2}V_{r}}{\partial r^{2}}+\frac{1}{r^{2}}\frac{\partial ^{2}V_{r}}{\partial\varphi ^{2}}+\frac{1}{r}\frac{\partial V_{r}}{\partial r}-\frac{2}{r^{2}}%
\frac{\partial V_{\varphi }}{\partial \varphi }-\frac{V_{r}}{r^{2}}\right) } & & (A2a)
\end{eqnarray}

\begin{eqnarray}\nonumber
{\frac{\partial V_{\varphi }}{\partial t}+V_{r}\frac{\partial V_{\varphi }}{%
\partial r}} {+\frac{V_{\varphi }}{r}\frac{\partial V_{\varphi }}{\partial
\varphi }-\frac{V_{r}V_{\varphi }}{r}} = {-\frac{1}{\rho _{0}r}\frac{\partial p}{\partial \varphi }+\frac{\partial r^{2}}{\partial r}\Omega \frac{V_{r}}{r}-}\Omega \frac{\partial V_{\varphi }%
}{\partial \varphi }+ \\ \nonumber
+{\nu \left( \frac{\partial ^{2}V_{\varphi }}{\partial
r^{2}}+\frac{1}{r^{2}}\frac{\partial ^{2}V_{\varphi }}{\partial \varphi ^{2}}%
+\frac{1}{r}\frac{\partial V_{\varphi }}{\partial r}-\frac{2}{r^{2}}\frac{%
\partial V_{r}}{\partial \varphi }-\frac{V_{\varphi }}{r^{2}}\right) } & & (A2b)
\end{eqnarray}

\begin{eqnarray}\nonumber
{\frac{\partial V_{z}}{\partial t}+V_{r}\frac{\partial V_{z}}{\partial r}+%
\frac{V_{\varphi }}{r}\frac{\partial V_{z}}{\partial \varphi }-V_{z}\frac{%
\partial V_z}{\partial z}} = {-\frac{1}{\rho _0}\frac{\partial p}{\partial z} }+ \\ \nonumber
+{\nu \left( \frac{\partial ^{2}V_{z}}{\partial r^{2}}+\frac{1}{r^{2}}\frac{%
\partial ^{2}V_{z}}{\partial \varphi ^{2}}+  
\frac{1}{r}\frac{\partial V_{z}}{%
\partial r}-\frac{2}{r^{2}}\frac{\partial V_{r}}{\partial z}-\frac{V_{z}}
{r^{2}}\right) } & & (A2c)
\end{eqnarray}

Because of the time averaging and the presence of perturbations in the
quantities, some of the terms in the equations may vanish or reform their
expression. The expression in z-direction brings no new information and
after averaging over z, Eq. (A2c) can be ignored. The second and third term
in the right hand site of Eqs. A2a and A2b, express turbulence activity.
This term would allow detection of sharp changes in the instability and is
called: "term of instability" (further shortly call: "inst term"). Now,
following the perturbed equations (A2a, A2b), and the effect of perturbation
function, next form of this "term" have been extracted: $\Omega \left(\frac{%
\partial r^{2} }{\partial r} \frac{V_{r} }{r} -\frac{\partial V_{\varphi } }{%
\partial \varphi } \right)$ and $2\Omega \left(\frac{1}{2} \frac{\partial
V_{r} }{\partial \varphi } -V_{\varphi } \right)$. They contain the quantity
of angular velocity $\Omega $ that is found to be of importance for studying
the differential flow in accretion discs. It appears that the "inst term" in
the above equation is significant only in $r$ and $\varphi $ directions, as
should be expected in accordance with the initial conditions for this case.
Then, using the mode number $m$ from the perturbation expression we examine
how the velocity and density variations develop. For the graphical
visualization of the disturbances effects, we use a grid-scale measurement
of values [12, 12, 12], which is more suitable to detect the results. The
next initial values are applyed: $V\left(0\right)=V_{0} $, $u\left(t_{0}
\right)=u_{0} $, $t_{0} \approx 1$, $r_{0} \approx 1$. It results in
occurrence of velocity excess during the period of disturbance (Fig. 5).

\begin{figure}[htp]
\centering\includegraphics[width=10cm, height=10cm]{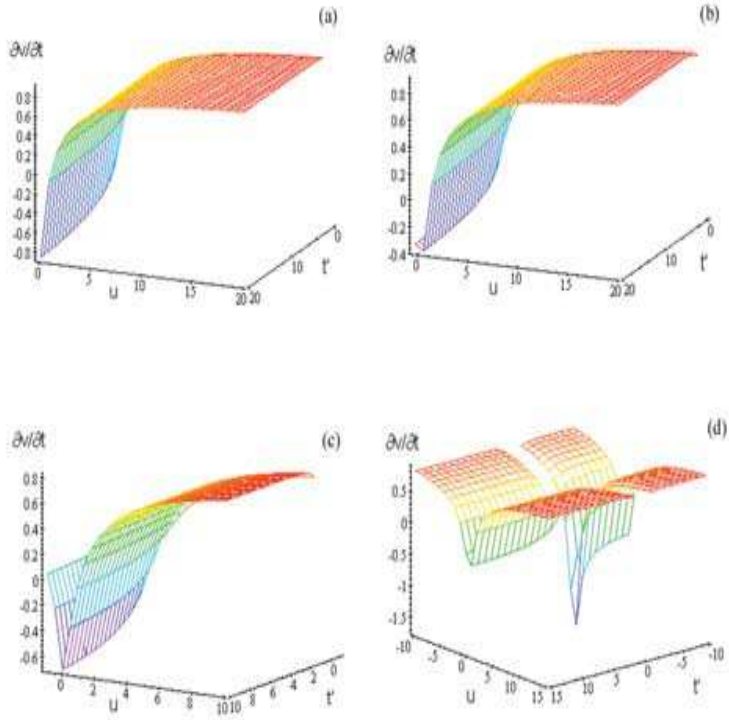}
\caption{Velocity variations in the flow during the mass transfer. It is
seen four consecutive phases that trace out the process in different values
of m (Fig.5(a),5(b),5(c),5(d)). When $m\in \left(-0.2\div -0.45\right)$ the
velocity state is in a quiet (Fig. 5a). Variations start at $-0.45>m\le -1$
(Fig. 5b, 5c). The critical value in Fig. 5(d) is at $m=-10$, when some
fission and sharply decrease of the velocity are observed.}
\label{Figure A.1}

\end{figure}

Figures 5(a-d) show how the process changes when different values of $m$
are applied. It is observed that in a range of $\left(-0.2\div -0.45\right)$
there is no velocity exchange. When the values of $m$ start to decrease ( $%
m\le -1$) the process reverses. This leads to a fission in the solution and
a suddenly drop-off in the velocity values is observed. This could be caused
by some local unstable activity in the accretion discs zone (e.g. the change
of mass transfer rate of the inflow matter in close binary, studied in
Sect.3). The scales of the axes are constrained by the parametrization of
the code.

In the next figure we show the result of applying the perturbation analysis
(explained above) on the matter of interaction flow after variation in mass
transfer rate. The calculations are made only for one orbital period of the
rotating system. Figure 6 shows the appearing of dense pattern, called
``thickened zone''.

\begin{figure*}[htp]
\centering\includegraphics[width=7cm, height=7cm]{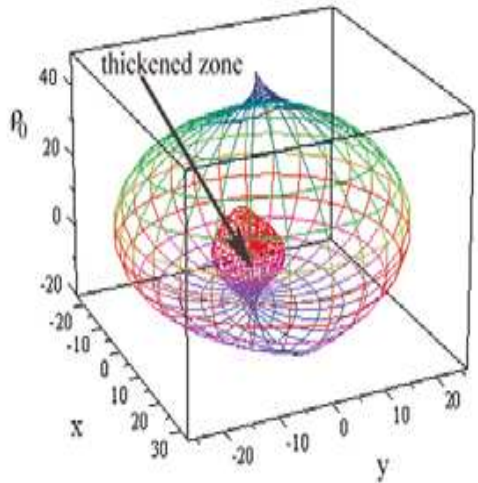}
\caption{Thickened zone formation in the flow as a result of disturbances in
the stability state, caused by mass transfer in the binary system.
Calculations are for one period of time with grid [30,30,30]. The place of dense formation is shown with arrow.}
\label{Figure A.2}
\end{figure*}

The analysis of the perturbation functions and the expression of
Navier-Stokes equations presented in this Appendix help us  to understand
the behavior of interacting flows in close binary stars, as the processes
studied in Sect.3 and 4 in the current paper.

\subsection*{Bibliography}

     Autar, K.K., Egwu, E.K. 2008, Numerical methods with applications, 1st ed., self-publ.,  
    
     Balbus, S. A. 2003, ARA\&A, 41, 555-597
 
    Balbus, S. A., Hawley, J. F.
1998, Rev. of Mod. Phys, 70, 1
   
   Barranco, J. A. Marcus, P. S. 2005, ApJ, 623, 1157
   
  Bisikalo D. V., Boyarchuk A. A.,
Kilpio A.A., Kuznetsov O. A., 2001, Astron. Rep., 45, 676
  
   Bisikalo, D.V., Boyarchuk, A.A., Kaigorodov
P.V., Kuznetsov O.A. 2003, Astron. Rep. 47, 809
  
  Boneva, D.V. 2009, Bg AJ, 11, pp. 53–65
  
  Boneva, D.V., Filipov, L.G. 2009, Proceedings
of SENS’09 (International conference of Space Ecology Nanotechnology and
Safety), Sofia, Bulgaria, p.360
  
    Boneva, D.V. 2010, BgAJ, 13, pp. 3-11, ISSN 1313-2709 (1)
  
    Boneva, D. 2010, AIP Conf. Proceedings, 1273,324
(2)
  
 Boyarchuk, A. A., Bisikalo,
D. V., Kuznetsov, O. A., Chechetkin V. M. 2002, Adv. in Astron. and
Astroph., Vol. 6, London: Taylor \& Francis
  
    Bracco, A., Provenzale, A., Spiegel, E., Yecko, P. 1998, in Theory of Black Hole Accretion Disks, ed. M. Abramowicz, G. Bjornsson, J. Pringle (Cambridge: Cambridge University Press), 254
  
    Chang, M.J., Chow, L.C., Chang, W.S. 1991, Numerical Heat transfer,
Part B, 19(1), 69-84, ISSN 1040-7790
  
    Clark, C., Carswell, R. 2007, Principles in Astrophysical Fluid Dynamics,
Cambridge University Press, ISBN-13: 978-0-511-27379-7
  
    Fridman, A.M. 2007, Phys. Usp., 50, 115
  
    Frank, J., King, A., Raine, D. 2002, Accretion Power in
Astrophysics, 3-rd edition, Cambridge University Press, New York 
  
   Frank, J. 2008, New Astronomy Reviews, 51,
878-883
  
  Felder R. M., Introduction to Maple, II. Elementary Calculus and Differential Equations, North Carolina State University,
  
   Godon, P. 1997, ApJ, 480, 329
  
   Godon, P., Livio, M. 1999, ApJ, 523,
350
  
   Godon, P., Livio, M. 2000, ApJ, 537, 396
  
   Gonzalez, R., Woods, R. 1992, Digital Image Processing,Addison-Wesley Publishing Company,  p 191.
  
   Graham, J.R. 2001, "Astronomy 202:
Astrophysical Gas Dynamics". Astronomy Department, UC Berkeley
  
  Hairer, E., Soderling G. 2005, SIAM J. Sci.
Comput., Vol. 26, 6, pp. 1838-1851

    Haddad, R.A., Akansu, A.N.
1991,IEEE Transactions on Acoustics, vol. 39, pp 723-727
  
    Heal, K.M., Hanse, M.L. Rickard, K.M. 1998, Maple V:
Learning Guide (for Release 5), Springer, New York.
  
    Johnson, B. M., Gammie, C. F. 2005, ApJ, 635,
149-156
    Johnson, B. M.,
Gammie, C. F. 2006, ApJ, 636, 63
  
  Klahr,  H., Bodenheimer, P. 2000,
Proceedings of: Discs, Planetesimals and
Planets, (Astronomical Soc. Of the
Pacific), vol. 219, 63
  
    Klahr, 
H., Bodenheimer P. 2003, ApJ, 582,
869-892 
   Klahr, H. 2004, ApJ, 606, 1070
  
   Lesur, G., Papaloizou, J.C.B.
2009, A\&A, manuscript no. 1557
  
    Li, H., Colgate,
S., Wendroﬀ, B., Liska, R. 2001, ApJ, 551, 874
    Li, H., Finn, J., Lovelace, R., Colgate, S. 2000, ApJ, 533, 1023
  
    Lithwick, Y. 2007, ApJ, 670, 789L
  
    Lithwick, Y. 2009, Ap J, 693, 85

    Lovelace, R.V.E., Li H., Colgate,
S.A.,
Nelson, A.F. 1999, Ap.J., 513, 805-810
    Monnagan, M.B., Geddes, K.O., Heal, K.M., Labhan, G., Vorkoetter,
S.M. 1998, Maple V: Programming Guide, 2nd edn. (for Maple V Realease 5),
Springer, New York
    Nauta,. D. 2000,
Two-dimensional vortices and accretion disks, University
Utrecht
  
   Nixon, M. S., Aguado, A.S. 2008,
Feature Extraction and Image Processing, Academic Press,  p. 88.
  
    Petersen, M.R., Steward,
G.R., Julien K.
2007, Ap J., 658:1252-1263 
  
    Papaloizou, J. C. B., Pringle, J. E. 1984, MNRAS,
208, 721-750,

  Pringle, J.E., 1985, in
Interacting Binary Systems (Chapter 1), eds. J.E. Pringle, R.A. Wade,
Cambridge:Cambridge University Press 
   
     Pringle,
J.E. 1992, A. Note, ASP Conf. Series, 22, 14
 
    Ritter, H. 1996, in Evolutionary Processes in Binary Stars,
NATO Advanced Science Institutes (ASI) Series C, Mathematical and
Physical
Sciences, Vol.  477, p. 223.

Rozyczka, M., Spruit, H.C. 1993, Ap J, 417, 677
 
    Sawada, K., Matsuda T., Hachisu I. 1986,
NRAS, 219, 75

  Shapiro, L.
G., Stockman, G. C. 2001, Computer Vision, page 137, 150, Prentence Hall 

    Shen, Y., Stone, J. M., Gardiner, T.
A.
2006, ApJ, 653, 513
     Steeghs,
D.,
Harlaftis, E.T., Horne, K. 1997, MNRAS, 290, L28.
    Shore, N.S. 2007, Astrophysical Hydrodynamics, 2-nd ed.,
WILEY-VCH
Verlag GmbH \& Co. KGaA, ISBN: 978-3-527-40669-2
    
Shu, F.H., (1992), The Physics of Astrophysics, Vol II: Gas
Dynamics,
University Science Books.  
    
Thorne, K.
2004, Foundations of fluid dynamics, V
0415.2.K2004,
http://www.pma.caltech.edu/Courses/ph136/yr2004/
   
Vandevenne, L., Lode's Computer Graphics
Tutorial Image Filtering, Copyright (c) 2004-2007,
http://lodev.org/cgtutor/filtering.html/Convolution

\end{document}